\documentclass[a4paper]{EslabStyle}
\usepackage{epsfig}

\begin{document}

\title{Combined color indexes and photometric
structure of galaxies \object{NGC 834} and \object{NGC 1134}}

\author{D.Bizyaev\inst{1} \and A.\,A.Zasov\inst{1} \and S.Kajsin
\inst{2}}
\institute{Sternberg Astronomical Institute -- Universitetskiy
prospect 13 Moscow, Russia  \and Special Astrophysical Observatory
of Russian Academy of Sciences -- Nizhniy Arkhyz, Karachaevo-Cherkessia,
Russia}

\maketitle

\begin{abstract}

We present the results of BVRI photometry of two galaxies with
active star formation: \object{NGC 834} and \object{NGC 1134}. 
Combined color index $Q_{BVI}$ was used to investigate 
the photometrical structure of the galaxies. Index $Q_{BVI}$ 
is not affected by internal extinction and is sensitive to 
the presence of blue stars.

Ring-like region with active star formation at 15"
from the center reveals itself in the $Q_{BVI}$ map of \object{NGC 834}.
Three-arm spiral structure is well-seen on the $Q_{BVI}$ map of NGC
1134.

We propose to use the combined indexes $Q_{BVI}$ and similarly
defined indices as a tracers of Star Formation activity and structure
of dusty galaxies.

\end{abstract}

\section{Introduction}

Observations of galaxies in optical wavebands are strongly influenced
by selective extinction which is hard to take into account due to
inhomogeneous distribution of dust. By this reason the maps of
color and brightness may give distorted picture
of photometrical structure of galaxies and distribution of
star-formation tracers.

In paper \cite*{zm98} it was proposed to use 
the combined color index $Q_{VRI}$
whose value doesn't depend on selective attennuation of light to
trace the structure of galaxies. In general case for photometric
bands A,B,C one can define combined index $Q_{ABC}$ as

$$ Q_{ABC} = (A-B) - \frac{E_{AB}}{E_{BC}}~ (B-C) $$

\noindent to make it as far insensitive to extinction as possible.
Here $\frac{E_{AB}}{E_{BC}}$ is the ratio of color excesses.  We have
assumed its values for VBRI system to be equal to standart values
for our Galaxy according to \cite*{c_89}: $\frac{E_{BV}}{E_{VI}}$ = 0.840,
$\frac{E_{VR}}{E_{VI}}$ = 0.413, $\frac{E_{BV}}{E_{VR}}$ = 2.033,
$\frac{E_{BR}}{E_{BI}}$ = 0.681. As it was shown in \cite*{zm98}, these
ratioes do not depend practically on whether we observe the source
through the dust screen or dust and stars are well mixed.
Similar approach was later used in \cite*{v99} to investigate
the structure of M 51.

\section{Observations}

Two spiral galaxies - \object{NGC 834} and \object{NGC 1134} 
were observed in BVRI
colors (Cousins system) 21-22 Jan 1988 at 1-m reflector of Special
Astrophysical Observatory of Russian Academy of Science. CCD camera
512x512 with scale 0."37/pix was used. Data processing was carried
out with the help of MIDAS package.

Asymmetric distribution of colors on the color maps 
of \object{NGC 834} and \object{NGC 1134}
gives evidence that they are dusty objects.

\section{Data analysis}

Combined colors indices do not describe neither "real"
colors nor the value of the extinction. They may be considered as
some conditional color indices free (or nearly free) of color
excesses.  The values of $Q_{ABC}$ depend on stellar population,
stellar abundances and, if R band is involved, on the presence of
$H_\alpha$ emission (see \cite{zm98}).

As an illustration, Fig.1a shows how the values of $Q_{BVR}$,
$Q_{VRI}$, $Q_{BVI}$ and $Q_{BRI}$ change when young stars formed
in the single burst with the age t=0.1 Gyr are overlapped onto the
the old population (t=13 Gyr) for different relative mass of young
stars.  The evolution program by \\ G.Worthey (see \cite{w94}) was
used adopting Salpeter IMF and [Fe/H]=0. Fig.1b
illustrates the dependence of Q's on the burst ages for a fixed (1$\%$)
fraction of young stars. Fig 1c shows a variation of Q's with the
adopted metallicity of the old population for the same parameters of
burst as in Fig 1b.


\begin{figure}[ht]
  \begin{center}
    \epsfig{file=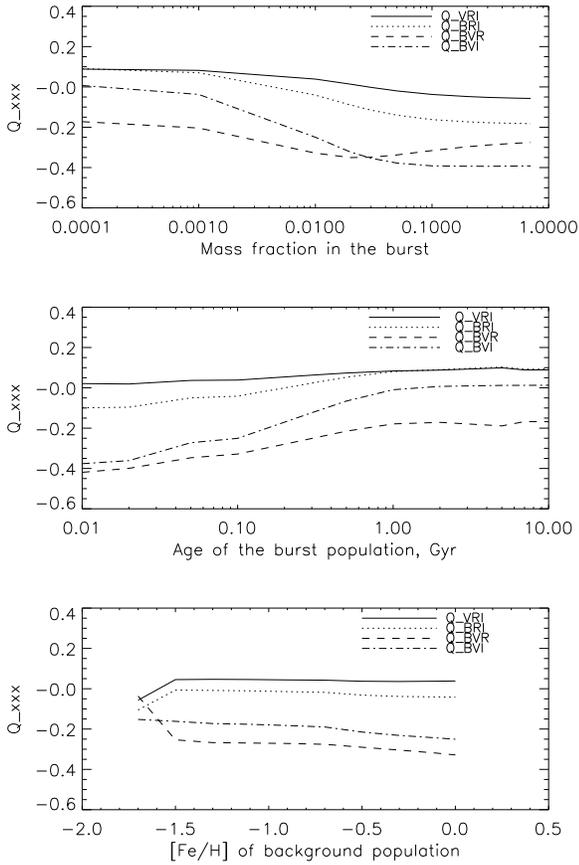, width=8cm}
  \end{center}
\caption{{\bf (a)} Variation of model values of
$Q_{BVR}$, $Q_{VRI}$, $Q_{BVI}$ and
$Q_{BRI}$ with relative mass of young population. Young stars
formed in the single burst with the age t=0.1 Gyr are overlapped
onto the the old population (t=13 Gyr).  The evolution engine by
G.Worthey was used adopting Salpeter IMF and [Fe/H]=0.
~~~{\bf (b)}Dependence of Q's on the burst ages for the fixed (1$\%$)
fraction of young stars.
~~~{\bf (c)}Variation of Q's with the adopted metallicity of the old
population for the same parameters of burst as in Fig 1b (burst age
0.1 Gyr).
\label{fig1}}
\end{figure}

As it follows from Figures 1a - 1c, $Q_{BVI}$ is the most sensitive
to the fraction of young stars. The change of $Q_{VRI}$ is less
significant, but on the other hand this index should allow to
localize giant emission regions where $H_{\alpha}$ line locally
increases R luminosity of the disk.

In Fig.2 and 3 the maps of $Q_{BVI}$ and $Q_{VRI}$ distribution for our
galaxies are presented.  Maps are bounded by the ellipses whose major axes
equal to $D_{25}$.  They appear to be much more symmetrically distributed than
color indices. Curiously a ring-like zone at about 15" (4.7 kpc for $H_0 = 75
~km s^{-1} Mpc^{-1}$) from the center appears in \object{NGC 834} which is not
noticeable at the color maps.  It enables to consider this system as the
galaxy, where star formation is enhanced in the ring. 
Note that \object{NGC 834} is
absent in catalog of a ring galaxies by \cite{bc93} ).

$Q_{VRI}$ map of \object{NGC 1134} reveals a clear 3-armed 
spiral-like structure
where the values of $Q_{VRI}$ have local maxima which may be explained by
enhanced emission in $H_{\alpha}$ (local equivalent width
$W_{H\alpha} \approx$ 100 $\AA$). However blue stars do not form a
clear spiral structure in this region as one can see from $Q_{BVI}$
map of the galaxy.


\begin{figure}[ht]
  \begin{center}
    \epsfig{file=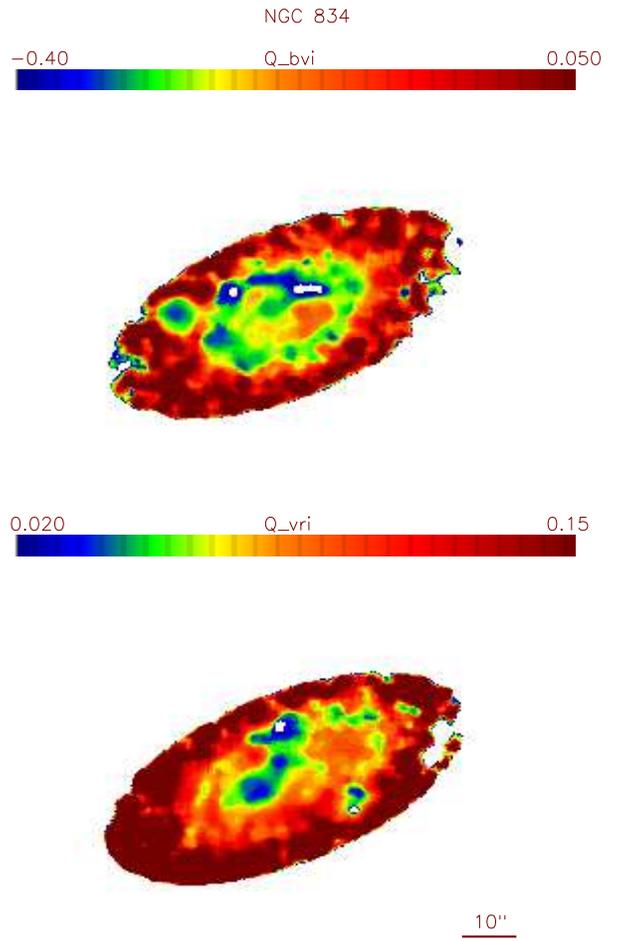, width=7.5cm}
  \end{center}
\caption{
$Q_{BVI}$ and $Q_{VRI}$ maps for \object{NGC 834}. They appear to be
much more symmetrically distributed than color indices.  The
ring-like zone at about 4.7 kpc from the center appears on $Q_{BVI}$
map.
\label{fig2}}
\end{figure}


\begin{figure}[ht]
  \begin{center}
    \epsfig{file=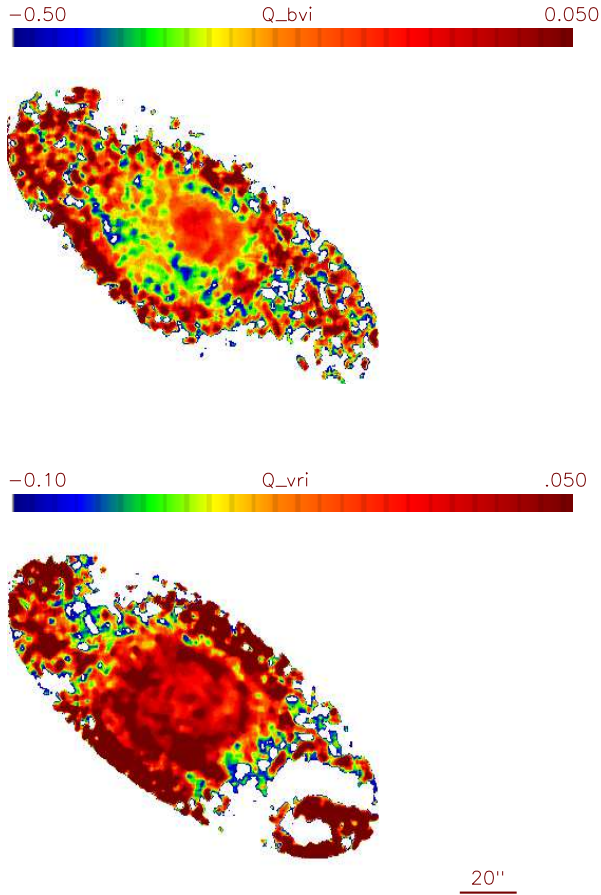, width=7.5cm}
  \end{center}
\caption{
$Q_{BVI}$ and $Q_{VRI}$ maps for \object{NGC 1134}.  $Q_{VRI}$ map of 
\object{NGC 1134} reveals 3-armed spiral-like structure where the values of 
$Q_{VRI}$ have local maxima which may be explained by enhanced emission in 
$H_{\alpha}$.
\label{fig3}}
\end{figure}

We can compare the relative intensity of star formation in
different regions of a galaxy using $Q_{BVI}$ -- $Q_{VRI}$ diagram.
The toy-model of a galaxy with exponential decreasing $SFR \sim
exp(-t/\tau)$ was developed using Worthey's program (from \cite{w94}).
The model
values of Q's which do not take into account recently formed stars
(t $<$ $10^9$ yr) are shown by continious curve in Fig.4.

Dashed curve shows the values of the combined indices
when the burst of SF is added to the previous model
(young population contains 30$\%$ of the mass of the background
population, has solar metallicity and the age of 0.1 Gyr).
The arrow in Fig.4. shows how $H\alpha$ emission shifts the point in
the diagramm.

The model curves show that the increasing of the fraction of
young stellar population change mostly $Q_{BVI}$. On the other hand
the $H_a$ shifts points horizontally. It enables to use
the diagram as the diagnostic one for comparizon of
star formation in different parts of a galaxy.

As an example we put average values of Q's for more than hundred
points of central region (diamond), regions of SF-ring
(triangle) and outer regions (square) of \object{NGC 834} 
on the diagram in Fig.4.


\begin{figure}[ht]
  \begin{center}
    \epsfig{file=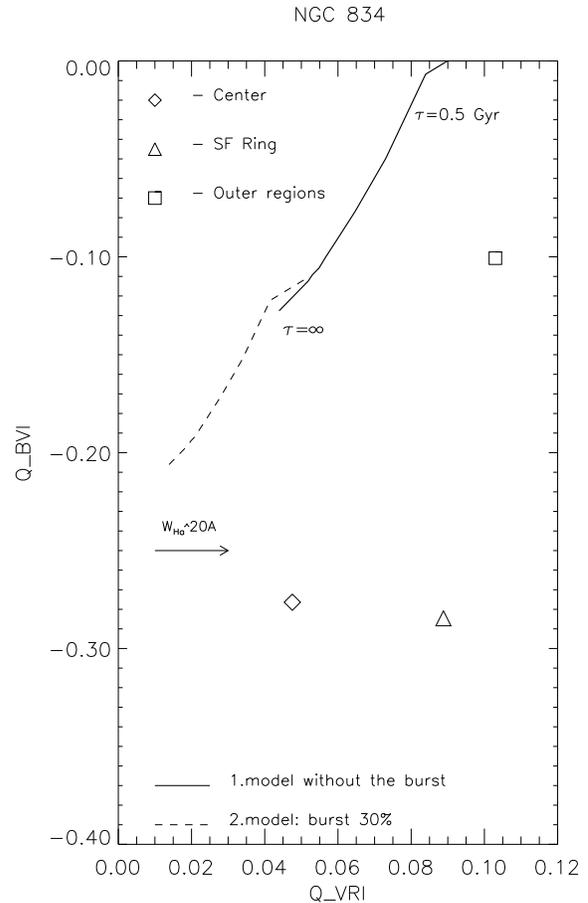, width=8cm}
  \end{center}
\caption{
Diagram $Q_{BVI}$ -- $Q_{VRI}$. Average value of Q's for the points of
central region of \object{NGC 834} is shown
by the diamond, value for the regions on the SF-ring are shown by the
triangle and values for the outer regions are denoted by the square.
~~~{\bf Curves:}
~~{\bf 1}. The toy-model of galaxy with decreasing $SFR
\sim exp(-t/\tau)$ was computed for different values of $\tau$. There
are no recent bursts.
~~{\bf 2}. Dashed curve shows the values of the combined indices
when the burst of SF is added to the previous model.
\label{fig4}}
\end{figure}

Mean errors of position of the points is about $0^m.003$.
Regions in the center and in the SF-ring have the same $Q_{BVI}$ which
enables to conclude that they have similar fraction of young stars.
The difference in $Q_{VRI}$ may be explained by the presence of $Ha$
emission (local $W_{H\alpha} ~\sim~ 50 \AA$). The outer regions
of the galaxy contain a less amount of blue stars.

These examples illustrate the ability to investigate
structure and distribution of star formation tracers in dusty
galaxies using optical broad-band photometry.

\section{Conclusion}

Combined photometrical indices $Q_{BVI}$ and $Q_{VRI}$ which are
weakly affected by the selective extinction may be used successfully
to restore the photometrical structure of galaxies with
non-homogeneous dust distribution, even if their dust-free colors
remain unknown.  Index $Q_{BVI}$ depends on the presence of blue
stars which enables to use it to localize regions of recent star
formation. On the other hand, $Q_{VRI}$ weakly depends on the light
of blue stars but is sensitive to young star population through the
emission in $H_\alpha$ line. Whereas color distribution is asymmetric
in \object{NGC 834} and \object{NGC 1134} due to internal 
absorption, their $Q_{BVI}$
and $Q_{VRI}$ maps are relatively symmetric and allow to reveal
structure details hidden by the dust. In \object{NGC 834} a ring-like region
evidently related to active star formation is clearly visible at
Q-maps.  In \object{NGC 1134} three armed spiral structure and active star
formation reveal themselves in $Q_{VRI}$ map of the galaxy.

This research  was suported by russian grants RFBR 98-02-17102 and Federal
program ''Astronomy''. D.B. very appreciates financial support of 
European Space Agency to participate to the 33rd ESLAB Symposium.

\end{document}